\documentclass[prl,aps,twocolumn]{revtex4} 
\usepackage{graphicx} 
\begin{document} 
\title{Free surface flows with large slopes: beyond lubrication theory}  

\author{Jacco H. Snoeijer}  
\affiliation{Physique et M\'ecanique des Milieux H\'et\'erog\`enes 
and  Mati\`ere et Syst\`emes Complexes, UMR 7636 and 7615 CNRS-ESPCI, 
10 rue Vauquelin, 75005, Paris, France}

\date{\today} 

\begin{abstract} 
The description of free surface flows can often be simplified 
to thin film (or lubrication) equations, 
when the slopes of the liquid-gas interface are small. 
Here we present a long wavelength theory that remains fully 
quantitative for steep interface slopes, by expanding about 
Stokes flow in a wedge. 
For small capillary numbers, the variations of the interface 
slope are slow and can be treated perturbatively. 
This geometry occurs naturally for flows with contact lines: 
we quantify the difference with ordinary lubrication theory 
through a numerical example and analytically recover the full 
Cox-Voinov asymptotic solution. 
\end{abstract}

\maketitle
\begin{figure}[tbp] 
\includegraphics[width=7.0cm]{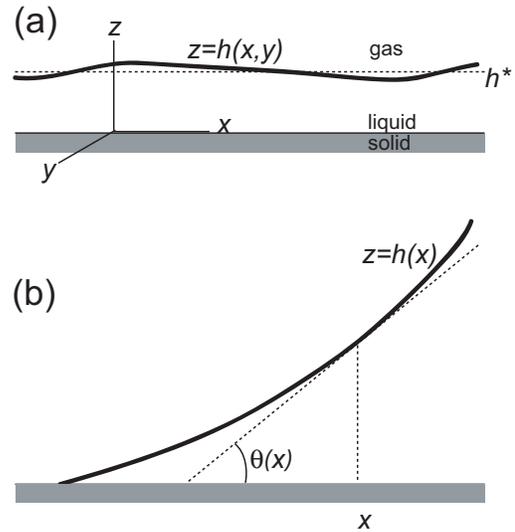} \centering 
\caption{(a) The usual lubrication approximation is valid whenever 
the liquid-gas interface is slowly varying along the horizontal 
coordinates, and is thus restricted to small slopes. 
(b) Considering wedge-like profiles with slowly varying slope $\theta(x)$, 
we derive the long wavelength theory for steep slopes.} 
\label{fig.schematic} 
\end{figure}

 
Free surface flows are encountered in many everyday life 
and industrial situations, ranging from soap films, 
sliding drops to paints and coatings \cite{oron,kistler}. 
The hydrodynamic description of these 'free boundary problems' still provides 
a challenge of great fundamental and technological interest. 
The difficulty lies in the intricate coupling between the liquid-gas 
interface and the flow inside the film, which gives rise to a broad 
variety of instabilities and interface morphologies 
\cite{oron,huppert,homsey,schatz,eggersreview,cohen}. 
In the case of 'thin' films, for which horizontal and vertical 
length scales are well separated (Fig.~\ref{fig.schematic}a), 
the problem is greatly reduced through a long 
wavelength expansion \cite{oron,reynolds,benny}. 
At low Reynolds numbers this so-called lubrication approximation 
yields a single nonlinear equation for the evolution of the interface 
profile $h(x,y,t)$, and forms the accepted theoretical framework for 
free surface flows. 
This reduction is possible whenever surface tension ($\gamma$) 
dominates over viscosity ($\eta$), so that the capillary number 
${\rm Ca}=\eta U^*/\gamma$ serves as a small parameter; 
$U^*$ denotes the velocity scale of the problem. 

The standard formulation of lubrication theory, however, has a 
severe drawback: it is only valid for small interface slopes, 
i.e. $|\nabla h |\ll 1$. 
While it is generally believed that the lubrication equation yields good 
qualitative predictions for larger slopes as well, 
one has to be careful with {\em quantitative} comparisons. 
This is particularly important for the problem of moving contact 
lines for which viscous forces tend to diverge as $h\rightarrow 0$ 
\cite{huhscriven,dussan}. 
The microscopic mechanisms that release this singularity are 
highly disputed 
\cite{voinov,cox,eggersslip,blakedeconinck,pismen}, 
and call for a fully quantitive description 
of experiments that often involve large contact angles. 
As it is practically infeasible to resolve the full hydrodynamic 
problem on all relevant length scales, ranging from molecular to millimetric, 
a simplified theory for finite slopes would be extremely valuable.

In this paper we present a generalization of the lubrication theory 
for free surface flows at low Reynolds numbers that remains exact for large slopes. 
The crucial observation is that in the limit of small ${\rm Ca}$, 
capillary driving requires slow variations of the interface curvature, 
but {\em there are no restrictions to the steepness of the interface}. 
We therefore consider the flow in a wedge with a finite, 
but slowly varying opening angle $\theta$ -- see Fig.~\ref{fig.schematic}b. 
This geometry naturally occurs for contact lines. 
Treating the variation as a perturbation around a straight wedge then 
yields the equation for the interface profile $h(x,y,t)$. 

This theory, summarized by Eqs.~(\ref{newequation},\ref{correction}), 
remains fully quantitative for large slopes when 
the curvature $\kappa \ll 1/h$, 
while it enjoys the same mathematical structure as the 
usual lubrication theory. 
We furthermore show that the equation reproduces the asymptotics for 
$\theta(x)$ as computed by Voinov \cite{voinov} and Cox \cite{cox}, 
in a relatively straightforward manner. 
However, the present work reaches beyond asymptotic relations: 
it describes all intermediate length scales as well, 
and allows to incorporate other forces such as a disjoining 
pressure or gravity.

\paragraph{Lubrication theory for free surface flows --} 
Before addressing the problem of finite slopes, 
let us first briefly revisit the lubrication approximation. 
In the limit of zero Reynolds number, the flow of incompressible 
Newtonian liquids is described by Stokes equations

\begin{eqnarray}
\nabla \cdot {\bf v} &=& 0, \label{divergence} \\
- \nabla p + \eta \Delta {\bf v}  - \nabla \Phi &=& {\bf 0} 
\quad \Longrightarrow \quad \nabla\times \Delta {\bf v} ={\bf 0}, \label{stokes}
\end{eqnarray}
where $p(x,y,z)$ and ${\bf v}(x,y,z)$ represent the 
pressure and velocity field respectively, 
while we consider body forces that derive from a potential $\Phi$. 
(Time dependent profiles will be discussed below.) 
The equation is complemented with the boundary condition 
of Laplace pressure at the free surface

\begin{equation}\label{laplace} 
p(z=h) =  p_0 - \gamma \kappa, 
\end{equation}
where $\kappa$ is the interface curvature. 
This gives rise to an intricate nonlinear coupling between the 
shape of the interface and the flow inside the film, 
which has to be resolved self-consistently. 
We limit the discussion to the case where the gas is hydrodynamically 
passive and we take a zero shear stress condition at the free interface.

In the limit where ${\rm Ca} \ll 1$, the free boundary 
problem can be greatly reduced through 
the lubrication approximation: surface tension is sufficiently strong to 
drive the viscous flow through only minor variations of the shape of the free surface 
(Fig.~\ref{fig.schematic}a). 
One thus expects that the interface profile and the velocity field are 
slow functions of the horizontal coordinates, so that 

\begin{eqnarray}\label{profileold}
\frac{h(x,y)}{h^*} &=& \tilde{h}\left(\epsilon \frac{x}{h^*}, \epsilon\frac{y}{h^*} \right), 
\end{eqnarray}
while we write

\begin{eqnarray}\label{velocityexpansion}
{\bf v}(x,y,z) &=& \tilde{{\bf v}}\left(\epsilon\frac{x}{h^*}, \epsilon\frac{y}{h^*},
\frac{z}{h^*} \right) \nonumber \\ 
&=&  \tilde{{\bf v}}_0 + \epsilon \, \tilde{{\bf v}}_1 + 
\epsilon^2 \,\tilde{{\bf v}}_2 + \cdots.
\end{eqnarray}
All lengths have been rescaled by the typical film thickness 
$h^*$, and $\epsilon={\rm Ca}^{1/3}$ is the small parameter of the expansion. 
The strategy is to solve Eqs.~(\ref{divergence}-\ref{laplace}) order by order in 
$\epsilon$. Here we briefly sketch the approach; 
for a more detailed derivation we refer to \cite{oron}. 

If we let $\epsilon$ go to zero, the departure of $h(x,y)$ from a 
horizontal interface becomes increasingly small. 
Hence, the velocity profile converges towards the parabolic 
(Poiseuille-like) profile in this limit, so that

\begin{eqnarray}\label{poiseuille}
\tilde{\bf v}_0 &=& \frac{3{\bf U}}{2} \left( 1 - 
\left(1-\frac{z}{h} \right)^2  \right).
\end{eqnarray}
Formally, this dominant flow can be obtained from Eq.~(\ref{stokes}) at order 
$\epsilon^0$. Note that $\tilde{\bf v}_0$ still evolves on a long scale 
through its dependence on $h(x,y)$. 

In deriving Eq.~(\ref{poiseuille}) we used the boundary conditions of no-slip at $z=0$, 
and zero shear stress at $z=h$. 
The prefactor has been chosen such that ${\bf U}\equiv 1/h \int_0^h dz \, {\bf v}_0$ 
represents the depth-averaged velocity in the frame 
attached to the plate. 

Since the dominant viscous forces in Eq.~(\ref{stokes}) arise from $\tilde{\bf v}_0$, 
we do not need to solve for the higher order velocities to obtain 
the equation for $h(x,y)$. 
At leading order, Eqs.~(\ref{stokes}) and (\ref{laplace}) 
reduce to the celebrated lubrication equation 

\begin{equation}\label{lubri}
\nabla \Delta h= 3{\rm Ca} \,\frac{{\bf U}/U^*}{h^2} +
\frac{1}{\gamma}
\nabla \Phi|_{z=h}.
\end{equation}
Combined with the depth-averaged continuity equation, 

\begin{equation}
\partial_t h + \nabla \cdot \left( h {\bf U} \right)=0,
\end{equation} 
it provides the common theoretical framework for free surface flows, 
both in the scientific community as well as for industrial purposes. 
The lowest order terms that are neglected are of order $\epsilon^2$, 
so the expansion is valid in the limit of small slopes, $|\nabla h|^2\ll 1$.

\paragraph{Theory for large slopes --}

We now perform a similar long wavelength expansion for wedge-like 
geometries, such as depicted in Fig.~\ref{fig.schematic}b. 
The crucial {\em physical} ingredient underlying the expansion is that 
surface tension dominates over viscosity, i.e. ${\rm Ca}\ll 1$, 
so that variations of interface {\em curvature} are slow with respect to the 
relevent length scales. 
In principle there is no restriction to the slope of the interface: 
the only requirement is that the opening angle $\theta$ is slowly varying 
(Fig.~\ref{fig.schematic}b). We therefore consider profiles with 

\begin{equation}
\theta(x) = \tilde{\theta}\left( \epsilon \frac{x}{x^*} \right),
\end{equation}
and perform an expansion in $\epsilon$. 
Note that $\tilde{\theta}$ itself is of order unity. 
Here we introduced the length scale $x^*$, which is a typical distance to 
the 'origin' of the wedge -- we show below that the analysis remains 
self-consistent as long as $h\, \partial_x \theta \ll 1$. 
For simplicity we discuss two-dimensional profiles so we omit 
the $y$-dependence. 

\begin{figure}[tbp] 
\includegraphics[width=5.5cm]{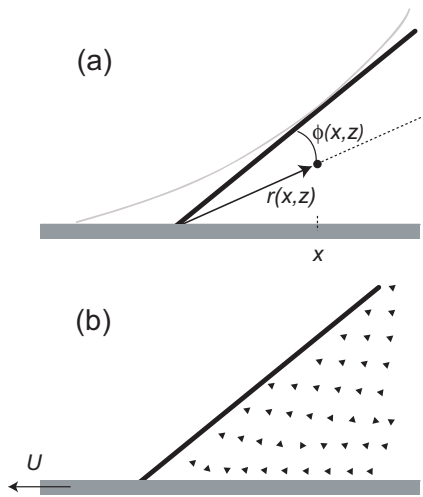} \centering 
\caption{(a) Definition of the cylindrical coordinates 
$r(x,z)$ and $\phi(x,z)$ in the locally tangent wedge of angle $\theta(x)$. 
(b) The basic velocity $\tilde{\bf v}_0$ corresponds to flow in a wedge of 
constant $\theta$, sketched in the frame comoving with the interface, 
Eq.~(\ref{wedgeflow}).} 
\label{fig.flow} 
\end{figure} 

In the spirit of the lubrication approximation, we expand the velocity as

\begin{equation}
{\bf v}(x,z) = \tilde{{\bf v}}_0 + \epsilon \, \tilde{{\bf v}}_1 + 
\epsilon^2 \,\tilde{{\bf v}}_2 + \cdots,
\end{equation}
and solve for the dominant flow $\tilde{{\bf v}}_0$. 
Again, $\tilde{\bf v}_0$ is obtained in the limit $\epsilon \rightarrow 0$, 
which in this case corresponds to a wedge of constant opening angle. 
Hence, the problem for $\tilde{\bf v}_0$ reduces to Stokes flow inside a 
straight wedge, which is easily solved analytically \cite{landau,huhscriven}. 

To make this more explicit, we introduce local cylindrical coordinates, 
$r(x,z)$ and $\phi(x,z)$, which are defined by the locally tangent wedge 
(Fig.~\ref{fig.flow}a)

\begin{eqnarray}
r(x,z) &=& \frac{h}{\tan \tilde{\theta}} \sqrt{1+
\left( \frac{z}{h} \tan \tilde{\theta}\right)^2 }, 
\nonumber \\
\phi (x,z) &=& \tilde{\theta} - \arctan \left( \frac{z}{h} \tan \tilde{\theta}  
\right). \nonumber
\end{eqnarray}
The $x$-dependence appears through $h(x)$ and $\theta(x)$. 
Writing the velocity as a function of these coordinates

\begin{eqnarray}
{\bf v}(x,z) &=& \tilde{{\bf v}}\left( r(x,z)\, , \, \phi(x,z) \right) 
\end{eqnarray}
and expanding Eqs.~(\ref{divergence},\ref{stokes}), 
one indeed finds that the order $\epsilon^0$ reduces to 
the problem of a straight wedge: variations of $\theta$ show up at higher orders. 
For the dominant flow we can thus use the results of \cite{landau,huhscriven}, 

\begin{eqnarray}\label{wedgeflow}
(\tilde{\bf v}_0)_r \! &=& U \,   
\frac{(\cos\phi  - \phi\sin\phi) \sin\theta - \theta\cos\theta\cos\phi}
{\theta -\cos \theta \sin\theta}, \nonumber \\
(\tilde{\bf v}_0)_\phi \! &=& U \, 
\frac{ \theta \sin \phi \cos\theta -  \phi\cos\phi \sin \theta  }
{\theta -\cos \theta \sin\theta},
\end{eqnarray}
which hold in the frame comoving with the interface. 
This flow has been sketched in Fig.~\ref{fig.flow}b. 
Here we used the conditions of a vanishing shear stress at $\phi=0$ 
and a no-slip condition at the plate, $v_r=-U$. 
The latter condition ensures that $U$ represents the 
depth-averaged velocity in the frame attached to the plate. 

Evaluating $\Delta {\bf v}_0$ at the free surface ($\phi=0$), 
Eq.~(\ref{stokes}) provides the leading order pressure gradients 
along the interface

\begin{equation}\label{pressure}
\partial_r p|_{\phi=0} = - \frac{2\eta U}{r^2}\left( 
\frac{\sin\theta}{\theta - \cos\theta \sin\theta}  \right) 
-\partial_r \Phi.
\end{equation}
Combined with the Laplace pressure condition (\ref{laplace}) 
this yields the generalized lubrication equation:

\begin{equation}\label{newequation}
\partial_x \kappa =  3{\rm Ca} \, \frac{U/U^*}{h^2} \, F(\theta) 
+ \frac{1}{\gamma}\, \partial_x \Phi|_{z=h},
\end{equation}
where 

\begin{equation}\label{correction}
F\left(\theta\right) =  \frac{2}{3} \,\frac{\tan \theta \sin^2 \theta}
{\theta - \cos \theta \sin \theta }. 
\end{equation}
Since $\partial_x \kappa \sim (\epsilon/x^*)^2$, the expansion is 
self-consistent when taking $\epsilon={\rm Ca}^{1/2}$. 

Comparing this result to the lubrication equation (\ref{lubri}), one observes two 
differences. First, the function $F(\theta)$ can be seen as a correction 
factor for the viscous term: 
indeed, one recovers $F(\theta) \rightarrow 1$ for small slopes. 
Second, the left hand side of Eq.~(\ref{newequation}) now involves the 
full curvature

\begin{equation}
\kappa = \frac{\partial_{xx} h}{(1+\partial_x h^2)^{3/2}}.
\end{equation}
The expansion thus provides an equation for the interface profile that 
has the same mathematical structure as the usual lubrication approximation, 
but which remains exact for large slopes. 
The lowest order terms that are neglected are of order 
$\epsilon = {\cal O}(x^* \partial_x \theta)$, 
so the expansion is valid when $h \partial_x \theta \ll 1$. 
This does not mean that the description is limited to small
heights: the height only appears as a reference scale to quantify the
dimensionless curvature.

\paragraph{Asymptotics for $\theta(x)$} -- 
To illustrate the strength of the approach, we now show that 
Eq.~(\ref{newequation}) correctly reproduces the nontrivial asymptotic 
solution for $\theta(x)$, 
as obtained by Voinov \cite{voinov} and Cox \cite{cox} for advancing 
contact lines. 
Anticipating the well known result, we express the relation between 
$\theta$ and $x$ as

\begin{equation}\label{bladiebla}
g(\theta) = {\rm Ca} \ln(x/x_0),
\end{equation}
and solve for $g(\theta)$ using Eq.~(\ref{newequation}) with $\Phi=0$. 
Differentiating $g(\theta)$ with respect to $x$ one finds 
$\partial_x \theta = {\rm Ca}/(x g')$, so that the curvature 
can be written as $\kappa= {\rm Ca}\, \cos \theta /(x\, g')$. 
Combining this with the expansion 
$h(x) =  x \tan \theta(x) + {\cal O}(\rm Ca)$, 
Eq.~(\ref{newequation}) becomes to lowest order

\begin{equation}
\frac{{\rm Ca} \, \cos \theta}{x^2 \, g'} = \frac{3{\rm Ca}}{x^2 \, \tan^2 \theta} F(\theta). 
\end{equation}
Here we used the fact that advancing contact lines move along the 
negative $x$-direction in the frame attached to the plate, so that $U_{\rm adv.}=-U^*$. 
Eliminating $g'$, we readily recover the famous result of \cite{voinov,cox}

\begin{equation}\label{cox}
g(\theta) = \int_0^\theta du \, \frac{u - \cos u \sin u}{2\sin u}.
\end{equation}

\begin{figure}[tbp] 
\includegraphics[width=8.0cm]{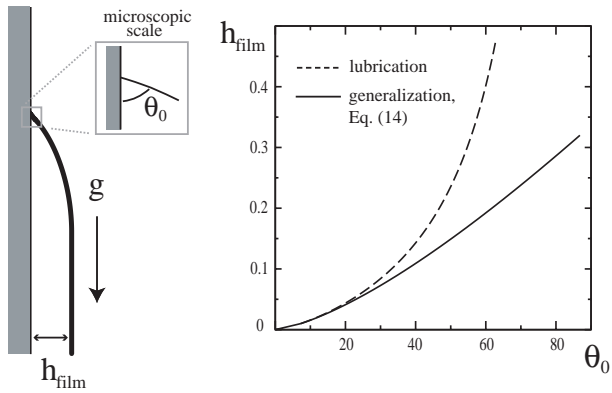} \centering 
\caption{Numerical solution of a flat film dragged downwards by gravity 
in the presence of a contact line. 
The film thickness $h_{\rm film}$ is uniquely determined by 
imposing a contact angle $\theta_0$ at a microscopic scale, 
here taken $h_0=10^{-5}$ (lengths are expressed in the capillary length 
$\sqrt{\gamma/\rho g}$). 
At large angles, the results of Eq.~(\ref{newequation}) (solid) 
provide significant corrections with respect to lubrication theory 
Eq.~(\ref{lubri}) (dashed).} 
\label{fig.numerics} 
\end{figure}

\paragraph{Discussion --} 
We have derived the long wavelength expansion for free surface flows 
in the case of steep interface profiles. 
This provides a significant improvement with respect to the usual 
lubrication approximation, whose validity is restricted to small slopes. 
The resulting theory has the same mathematical structure as the lubrication 
equation and is thus easily adapted to existing codes and methods. 
The most natural application of our work is found in wetting 
flows that involve large contact angles. 
This is illustrated in Fig.~\ref{fig.numerics}, where we computed 
the thickness of a flat film flowing down a vertical 
plate in the presence of a receding contact line. 
The effect of gravity is accounted for through $\Phi=-\rho g x$, 
which introduces the capillary length scale 
$l_\gamma = \sqrt{\gamma/\rho g}$. 
For a given thickness $h_{\rm film} = l_\gamma \sqrt{3{\rm Ca}}$, 
there is a unique solution that connects to the contact line \cite{hocking}. 
By numerically solving the interface profile down to a microscopic (molecular) 
height $h_0$, 
we can thus identify the slope very close to the contact line, 
denoted by the angle $\theta_0$ \cite{footangle}. 
This yields a unique relation between the slope 
imposed at a microscopic distance from the contact line and 
the macroscopic film thickness. 
The results obtained from numerical integration of 
Eq.~(\ref{newequation}) (solid) 
displays significant quantitative differences with the 
prediction by lubrication theory (dashed) 
at angles $\gtrsim 30^\circ$. 

Although in principle the expansion is not limited to contact lines, 
steep slopes are typically attained through curvatures $h h''\sim 1$ 
in flows without contact lines. 
This lies beyond the strict validity of the expansion and one expects 
corrections due to curvature of the interface. 
These corrections can in principle be treated perturbatively 
as well, since ${\rm Ca}\ll 1$ implies weak variations of curvature.  

It is a pleasure to thank Bruno Andreotti, Jens Eggers, 
Marc Fermigier, Alexander Morozov and Julien Tailleur. 
This work was supported by a Marie Curie Intra-European 
Fellowships (MEIF-CT2003-502006) within the 6th European 
Community Framework Programme.


\begin{thebibliography}{99} 

\small 

\bibitem{kistler} S.F. Kistler and P. Schweizer, 
{\em Liquid film coating - Scientific principles and 
their technological implications}, (Kluwer, Dordrecht, 1997).

\bibitem{oron} A. Oron, S.H. Davis and S.G. Bankoff, 
Long-scale evolution of thin liquid films, 
Rev. Mod. Phys. {\bf 69}, 931 (1997). 

\bibitem{huppert} H.E. Huppert, 
Flow and instability of a viscous gravity current 
down a slope, 
Nature {\bf 300}, 427 (1982).

\bibitem{homsey} R. Goodwin and G.M. Homsey, 
Viscous flow down a slope in the vicinity of a contact line, 
Phys. Fluids A {\bf 3}, 515 (1991).

\bibitem{schatz} N. Garnier, R.O. Grigoriev and M.F. Schatz, 
Optical manipulation of microscale fluid flow, 
Phys. Rev. Lett. {\bf 91}, 054501 (2003).

\bibitem{eggersreview} J. Eggers, 
Nonlinear dynamics and breakup of free-surface flows, 
Rev. Mod. Phys. {\bf 69}, 865 (1997).

\bibitem{cohen} I. Cohen and S.R. Nagel, 
Scaling at a selective withdrawal transition through a 
tube suspended above the fluid surface, 
Phys. Rev. Lett. {\bf 88}, 074501 (2002). 

\bibitem{reynolds} O. Reynolds, 
On the theory of lubrication and its application to 
Mr. Beauchamp Tower's experiments, 
including an experimental determination of the 
viscosity of olive oil, 
Philos. Trans. R. Soc. London {\bf 177}, 157 (1886).

\bibitem{benny} D.J. Benney, 
Long waves on liquid films, 
J. Math. Phys. (N.Y.) {\bf 45}, 150 (1966).

\bibitem{huhscriven} C. Huh and L.E. Shriven, 
Hydrodynamic model of steady movement of a solid/liquid/fluid contact line, 
J. Colloid Interface Sci. {\bf 35}, 85 (1971).

\bibitem{dussan} E.B. Dussan, V. Davis, and S.H. Davis, 
On the motion of a fluid-fluid interface along a 
solid surface, J. Fluid Mech. {\bf 65}, 71 (1974).

\bibitem{voinov} O.V. Voinov, 
Hydrodynamics of wetting, 
Fluid Dynamics, {\bf 11}, 714 (1976).

\bibitem{cox} R.G. Cox, 
The dynamics of the spreading of liquids on a solid surface, 
J. Fluid Mech. {\bf 168}, 169 (1986).

\bibitem{eggersslip} J. Eggers, 
Toward a description of contact line motion at higher capillary numbers, 
Phys. Fluids {\bf 16}, 3491 (2004).

\bibitem{blakedeconinck} T.D. Blake, J. De Coninck and 
U. D'Ortuna, Models of wetting: immiscible lattice boltzmann 
automata versus molecular kinetic theory, 
Langmuir {\bf 11}, 4588 (1995).

\bibitem{pismen} L.M. Pismen, Y. Pomeau, 
Disjoining potential and spreading of thin liquid layers in 
the diffuse-interface model coupled to hydrodynamics, 
Phys. Rev. E {\bf 62}, 2480 (2000).

\bibitem{landau} L.D. Landau and E.M. Lifschitz, {\em Fluid Mechanics}, 
(Pergamon, London, 1959).

\bibitem{hocking} L.M. Hocking, Meniscus draw-up and draining, 
Euro. Jnl. of Applied Mathematics {\bf 12}, 195 (2001). 

\bibitem{footangle} The problem of {\em predicting} the effective boundary 
condition $\theta_0$ at a microscopic scale $h_0$ lies at the heart 
of the contact line problem, and involves new microscopic mechanisms 
beyond classical hydrodynamics. 
The curves of Fig.~\ref{fig.numerics} depend only weakly 
(logarithmically, see Eq.~(\ref{bladiebla})) on the value of $h_0$.


\end{thebibliography}
\end{document}